\documentclass[pre,amsmath,amssymb,aps,floatfix,nofootinbib]{revtex4}

\usepackage{graphicx} \usepackage{amsmath} \usepackage{amssymb}

\usepackage{bm}

\usepackage{upgreek}

\usepackage{ulem}
\usepackage[switch]{lineno}

\newcommand{\comment}[1]{}

\newcommand{\BEA}{\begin{eqnarray}}
\newcommand{\EEA}{\end{eqnarray}}

\newcommand{\bq}{\begin{equation}}
\newcommand{\eq}{\end{equation}}
\newcommand{\be}{\begin{eqnarray}}
\newcommand{\ee}{\end{eqnarray}}
\newcommand{\ba}{\begin{align}}
\newcommand{\ea}{\end{align}}

\newcommand{\de}{\delta}
\newcommand{\om}{\omega}
\newcommand{\ep}{\epsilon}
\newcommand{\vep}{\varepsilon}
\newcommand{\bvep}{\eta}
\newcommand{\pa}{\partial}

\newcommand{\bx}{\bm{x}}
\newcommand{\br}{\bm{r}}
\renewcommand{\ba}{\bm{a}}
\newcommand{\by}{\bm{y}}

\newcommand{\bk}{\bm{k}}

\newcommand{\E}{F }

\begin{document}

\title{Resonances in light scattering from nonequilibrium dipole pairs}

\author{Vanik E. Mkrtchian$^{1)}$, Armen E. Allahverdyan$^{1)}$, Mikayel Khanbekyan$^{2)}$}
\affiliation{$^{1)}$Alikahanyan National Science Laboratory (Yerevan Physics Institute), 2 Alikhanyan Brothers Street, Yerevan 0036, Armenia\\
$^{2)}$Institute for Physical Research, Armenian National Academy of Sciences, Ashtarak-2, 0203, Armenia}

\begin{abstract}

We identify a fundamentally new class of resonances in light scattering from a pair of point-like electric dipoles. These resonances arise from multiple scattering response and correspond to poles of the two-particle Green function. The realization of exact resonances requires violation of the optical theorem, which holds in equilibrium systems but can be violated under nonequilibrium conditions, without violating causality. 
{As a concrete example, we show that a pumped two-level atom can realize the required polarizability and access the exact resonance regime. Our results for one electric and one magnetic dipole show how resonances can amplify a weak magnetic response of a single dipole to the incident field. In equilibrium systems, only approximate resonances with finite amplification are possible. Using the Drude model for metallic nanoparticles, we show that while equilibrium resonances enhance the scattering, they never exceed the optimal single-dipole response.}


\end{abstract}


\comment{Submission motivation:
We show the impact of a nonequilibrium medium on light scattering. The accounting for such effects requires multiple-scattering approaches. Hence, we study two point-like dipoles with monochromatic incident light, which allows us to study scattering without further approximations in classical optics. Nonequilibrium states are modeled by single dipole polarizabilities that do not satisfy the optical theorem. We show that in such dipole pairs, there are two-particle resonances that can be infinitely large both in near-field and far-field zones. Such resonances show up as exact poles of the two-particle Green function. In an approximate form (as approximate poles of the Green function), the resonances extend to the equilibrium domain, where the optical theorem holds. As an example of this, we show that the known plasmon resonance of a single equilibrium metallic nanoparticle can be enhanced more than 100 times via a pair of equilibrium nanoparticles. Even in the equilibrium domain, the effect is sufficiently strong to invalidate the Rayleigh limit of scattering. We envisage applications of the effect in building sensors. One possible application to enhancing magnetic features is explained here: the resonances between a nonequilibrium electric dipole and an equilibrium magnetic dipole amplify the small, single-particle magnetic response of the latter.
}

\maketitle


A wealth of resonance phenomena exists in light scattering \cite{novotny2012principles,klimov,bohren2008absorption,mishchenko2006multiple,mie}. They can be classified by their physical origin: {\it (i)} Morphological resonances (Mie, Fabry-P\'erot, Whispering Gallery, dielectric resonances) arise from the shape of the scatterer that creates standing wave patterns which cause the resonance. {\it (ii)} Materials have intrinsic resonances, e.g., plasmons, that are created by electrons driven by the incident electric field. {\it (iii)} Collective resonances, e.g., Fano, super-radiant modes, {\it etc}. Exceptional points also belong to {\it (iii)}. They arise from the coupling of two resonant modes in an open (non-Hermitian) system where eigenvalues and eigenvectors coalesce \cite{khanbekyan}. 

We consider light scattering from two point-like dipoles and demonstrate the existence of a fundamentally new class of collective resonances that arise from multiple scattering between the dipoles and refer to poles of the two-particle Green function. 
We report on resonances that share features with several known classes. Similar to {\it (i)} {they are} associated with a fixed geometric structure. However, they do not rely on any wave-confinement, even though constructive interference along closed optical paths still takes place. Instead, they utilize the near-field coupling between the dipoles. The resonances are also related to {\it (ii)}, since they are collective in nature, but they do not require mode hybridization or non-Hermitian structures; cf.~{\it (iii)}.

A key feature of the resonances described here is that their exact realization requires a nonequilibrium structure, that is a violation of the optical theorem (OT); cf.~{\it (iii)}. OT constrains passive systems, by imposing a lower {bound} on the imaginary part of the polarizability, relating it to the field extinction due to scattering \cite{markel1992scattering, wolf_ot}. However, this constraint does not follow from causality, and can be overcome in nonequilibrium systems, e.g., due to external pumping. The required polarizability can be achieved in active systems, e.g., a pumped two-level atom, enabling access to exact resonances.


Exact resonances correspond to
poles of the full scattering Green function. They are infinitely large within our setup, though achieving large values requires fine-tuned parameters. In the equilibrium regime, where the dipole polarizabilities hold OT, the resonances can still be observed, but they are finite and relate to approximate poles of the Green function. We exemplify this fact with the pair of gold nanoparticles described via the Drude model. Such a pair can amplify $\sim 10^2$ times the single-particle plasmonic resonance. Nevertheless, equilibrium resonances do not outperform the optimal single-particle response. The global amplification of the single-particle scattering is possible only via violation of OT. In a possible application of our result, we show that a nonequilibrium electric dipole, when paired with an equilibrium weak magnetic dipole, can enhance the magnetic response of the latter.


{\it Scattering from point-like dipoles.} Consider a monochromatic electromagnetic field ($k=1,2,3$):
\be
E_k(\bm{x},t)=e^{-i\om t}E_k(\bm{x},\om),~~ 
H_k(\bm{x},t)=e^{-i\om t}H_k(\bm{x},\om). 
\label{ii}
\ee
The two Maxwell's equations for 
$E_k(\bm{x},\om)=\E^1_k$ and $H_k(\bx,\om)=\E^2_k$ read in Gaussian units and in a representation that combines spinors (denoted by upper indices $a,b=1,2$) and 3-dimensional vectors (lower indices $i,j,k=1,2,3$) \cite{landau}: 
\be
\label{x2}
&M\begin{bmatrix}
\E^1_k\\
\E^2_k
\end{bmatrix}=
\eta
\begin{bmatrix}
\E^1_k\\
\E^2_k
\end{bmatrix},  ~~~~
M\equiv \begin{bmatrix}
\delta_{ik} & \frac{1}{i\om}\ep_{ijk}\pa_j \\
\frac{-1}{i\om}\ep_{ijk}\pa_j  & \delta_{ik} 
\end{bmatrix},
\\
&
\eta_{ik}^{ab}(\bx)=[1-\vep(\bx,\om)]\delta_{ik}\delta^{a1}\delta^{b1}
+[1-\mu(\bx,\om)]\delta_{ik}\delta^{a2}\delta^{b2},
\ee
where in formulas (but not in numerical estimates) we use $c=1$, we imply summation over repeated indices, $\delta_{ik}$ is the Kroenecker symbol, $\bx=(x_1,x_2,x_3)$, $\pa_i=\pa_{x_i}$, and $\ep_{ijk}$ is the fully antisymmetric tensor: $\ep_{123}=1=-\ep_{213}$. $\vep(\bx,\om)$ and $\mu(\bx,\om)$ are scalar space-dependent scalar electric permittivity and magnetic permeability, which characterize the scattering medium. The electromagnetic field in (\ref{x2}) reads
\be
\label{repre}
F(\bx)=F^{[\rm in]}(\bx)+F^{[\rm sc]}(\bx),\quad MF^{[\rm in]}(\bx)=0,
\ee
where the incident field $F^{[\rm in]}$ holds free equations (\ref{x2}), while $F^{[\rm sc]}$ is the scattered field. We focus on $n=1,2$ point dipole scatterers:
\be
\label{16}
\bvep(\bx)={\sum}_{\sigma=1}^n\bvep_\sigma\delta(\bx-\ba_\sigma),\qquad 
\bvep_\sigma=\{\bvep_{\sigma,\,ik}^{ab}\}, 
\ee
where $\quad i,k=1,2,3$; $a,b=1,2$, and $\bvep_\sigma$ is the electromagnetic polarizability of the dipole located at $\ba_\sigma$.
Eq.~(\ref{16}) implies that both the wavelength $1/\om$ and inter-particle distances 
$|\ba_\sigma-\ba_\mu|$ are much larger than the dipole particle's size:
\be
{1}/{\om},\,\,  |\ba_\sigma-\ba_\mu| \gg \ell\equiv {\rm dipole~size}.
\label{popo}
\ee
{For nanoparticles with $\ell\sim 10^{-6}$ cm and $\om\sim 10^{14}\,s^{-1}$, conditions (\ref{popo}) amount to $|\ba_\sigma-\ba_\mu|\gg 10^{-6}$ cm.

In (\ref{repre}), we assume} that $\E^{\rm [in]}$ is a plane wave 
$\E^{\rm [in]}(\ba_\sigma)=e^{i\bm{k}\ba_\sigma}\E^{\rm [in]}(0)$.
where $\bk$ is the wave-vector: $|\bk|=\om$.


{\it Two identical electric dipoles: exact resonances and OT.}
Now we have in (\ref{16}):
$\bvep_{1\, ik}^{ab}(\om)=\bvep_{2\, ik}^{ab}(\om)
=-\alpha(\om)\delta_{ik}\de^{a1}\de^{b1}$, 
where $\alpha(\om)/(4\pi)$ is the dipole polarizability. 
Section 1 of Supplementary material (SM) explains 
{how to get the following set of equations from
Eqs.~(\ref{ii}--\ref{16})}
\be
\label{333}
\begin{bmatrix}
E^{\rm [sc]}_k(\bx) \\
H^{\rm [sc]}_k(\bx)
\end{bmatrix}
= \frac{\hat\alpha e^{i\ba_1(\bk-\om\hat\bx)+i\om x}}{\om x} 
\begin{bmatrix}
(\de_{kl}-\hat x_k\hat x_l) \hat\Theta_{ln} E_n^{\rm [in]}(0) \\
-\ep_{kls}\hat x_s \hat\Theta_{ln} E_n^{\rm [in]}(0) \\
\end{bmatrix},
\label{kosh}
\ee
where $\hat \alpha\equiv \frac{\alpha\om^3}{4\pi}$ is the dimensionless one-particle polarizability, while $\hat\Theta_{ln}$ is the effective two-particle polarizability. It reads via dimensionless variables and via choosing the reference frame such that 
$\bm{r}=\ba_1-\ba_2=(r,0,0)$:
\begin{align}
\label{kotik}
&\hat\Theta_{11}=\frac{ 1+\xi\eta-(\xi+\eta)(\hat A+\hat B)}{ 1-(\hat A+\hat B)^2},~~
\xi=e^{-i\bm{k}\cdot\br},~~~ \zeta=e^{i\om\hat{\bm{x}}\cdot\br},\\
\label{motik}
&\hat\Theta_{22}=\hat\Theta_{33}
=\frac{ 1+\xi\zeta-(\xi+\zeta)\hat A}{ 1-\hat A^2},~~~~ 
\hat\Theta_{i\not=k}=0,\\
\label{on5}
&\hat g_{ik}=\hat A\delta_{ik}+\hat B\hat r_i\hat r_k, ~~~~
\hat A={\hat \alpha e^{iy}}y^{-3}(1-y(i+y)), ~~ \hat{\bm{r}}=\bm{r}/r, \\ 
&\hat B={\hat \alpha e^{iy}}y^{-3}(-3+y(3i+y)),~~~~
y=\om r,
\label{on6}
\end{align}
where $\hat\alpha$ is the dimensionless polarizability.
Hence, the resonances are defined via 
\be
\label{reso}
\hat A+\hat B=\pm 1~ {\rm and}~ \hat A=\pm 1.
\ee
{The physical meaning of these conditions is that they are the poles of the full Green function: they indicate on enhanced interparticle scattering; see section 1 of SM and Eqs.~(S15, S16). Conditions in (\ref{reso}) mean the scattered-field amplification in both the near- and far-field zones.  }
Eq.~(\ref{reso}) shows that in order to obtain 
$1+\hat A=0$, $\hat\alpha$ should satisfy: 
\BEA
\label{bruno3}
\hat \alpha=
{y^3 e^{-iy}}{(y^2+iy-1)^{-1}},\quad y=\om r.
\EEA
$\hat\alpha$ in (\ref{bruno3}) does not satisfy OT; see \cite{wolf_ot} for a review on this theorem. We shall need the limit of OT that applies to point dipoles: 
\BEA
{\rm Im}[\hat \alpha]\geq 2|\hat \alpha|^2/3.
\label{bruno5}
\EEA
The meaning of (\ref{bruno5}) is that for a passive (equilibrium) system ${\rm Im}[\hat \alpha]>0$ has to be sufficiently large, since it accommodates both the internal dissipation (e.g., heating), and the dissipation due to scattering. The form (\ref{bruno5}) of OT applies only to point (electric and/or magnetic) dipoles; cf.~(\ref{popo}). We do not need its more general forms, since we confine ourselves to point-like particles. But we still mention that for not point-like, passive systems, arbitrary small values of ${\rm Im}[\hat \alpha]>0$ are possible \cite{landau}.

\comment{
Precisely in the point-particle limit, the standard derivation of the optical theorem does not apply. Appendix shows how to deduce (\ref{bruno5}) starting from Poynting's theorem. 
}

\begin{figure}
\includegraphics[width=0.6\columnwidth]{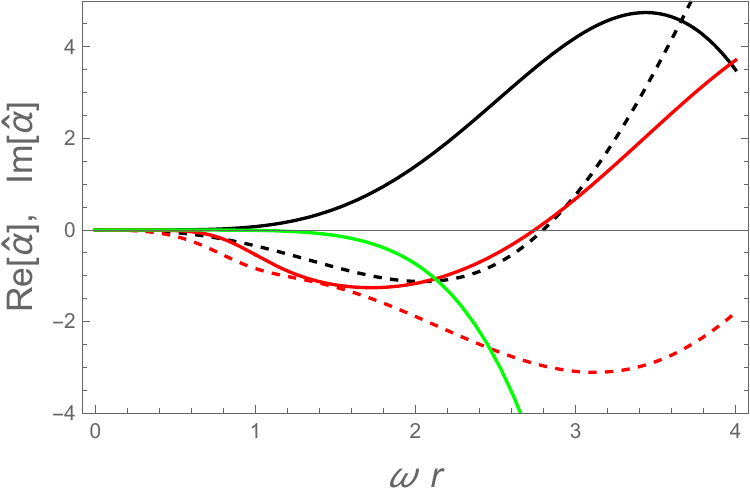}%
  \caption{The dimensionless polarizabilities ${\rm Im}[\hat\alpha]$ and 
  ${\rm Re}[\hat\alpha]$ at the exact resonance as a function of $\om r\equiv y$. Red (red dashed) curve: ${\rm Im}[\hat\alpha]$ (${\rm Re}[\hat\alpha]$) from (\ref{bruno3}). Green curve: ${\rm Im}[\hat\alpha]-\frac{2}{3}|\hat\alpha|^2$ calculated 
  from (\ref{bruno3}). The violation of OT (\ref{bruno5}) is tiny for $\om r\lesssim 1.5$.
  Black (black dashed) curve: ${\rm Im}[\hat\alpha]$ (${\rm Re}[\hat\alpha]$) 
  calculated for the exact resonance $\hat A+\hat B=1$, i.e., $\hat \alpha=\frac{1}{2}\,
\frac{y^3 e^{-iy}}{iy-1}$; see (\ref{reso}) and (\ref{on5}, \ref{on6}). Now ${\rm Im}[\hat\alpha]>0$ for considered $y=\om r$.  }
  \label{fig0}
\end{figure}

The poles $y_{\pm}=(-i\pm\sqrt{3})/2$ of (\ref{bruno3}) are in the lower complex semi-plane, i.e., (\ref{bruno3}) holds causality \cite{landau}. (Recall our frequency convention $\propto e^{-i\om t}$ in (\ref{ii}).) This means that (\ref{bruno3}) can, in principle, be realized via nonequilibrium systems. {All such systems hold causality, and no physical implementation, including artificial or metamaterial systems, can violate causality. Note that} ${\rm Re}[\hat \alpha]$ from (\ref{bruno3}) does not have a definite sign as a function of $y$. Likewise, ${\rm Im}[\hat \alpha(y)]$ changes its sign (oscillates) as a function of $y$. In particular, ${\rm Im}[\hat \alpha(y)]<0$ for $y\in (0.,2.75)$, and ${\rm Im}[\hat \alpha(y)]>0$ for $y\in (2.75, 6.1)$. Recall that ${\rm Im}[\hat\alpha]<0$ --  one mechanism of breaking OT -- is realized in lasers. 

{\it Exact resonances realized via pumped two-level system.}
The nonequilibrium polarizability needed for the exact resonances in (\ref{reso}) can be realized via a pumped two-level system (atom). Consider a two-level system with energy {level difference} $\hbar\omega_0$, dipole moment $\bm{d}$ and decay rate $\Gamma$, which is pumped by a strong field $\bm{E}=e^{-i\Omega t}\bm{E}_0$ at frequency $\Omega$. Then the response of the atom to the weak incident (probe) field at frequency $\om$ is described via a dimensionless polarizability $\hat\alpha$ \cite{boyd}:
\be
\label{bo}
&\hat\alpha=
-\frac{3 \left(4 \epsilon ^2+1\right) \left((\delta +i) \left(\delta -\epsilon +\frac{i}{2}\right)-\frac{\delta  | V| ^2}{2 \epsilon
   -i}\right)}{4 \left(2 | V| ^2+4 \epsilon ^2+1\right) \left[(\delta +i) \left(\delta -\epsilon +\frac{i}{2}\right)\, \left(\delta +\epsilon
   +\frac{i}{2}\right)-\left(\delta +\frac{i}{2}\right) | V| ^2\right]},
\ee
with dimensionless parameters
\be
   \epsilon=(\Omega-\om_0)/\Gamma, \quad
      \delta=(\omega-\Omega)/\Gamma, \quad
        V=|\bm{d}\cdot\bm{E}_0|/\Gamma.
\ee
For $V\to 0$ (no pump field), we get from (\ref{bo}):
\be
\label{bo2}
\hat\alpha=
-\frac{3}{4}\, \Big[\frac{\om-\om_0}{\Gamma}+\frac{i}{2}\Big]^{-1},
\ee
a known expression for the polarizability of the two-level atom \cite{dipole_many_2016}. Eq.~(\ref{bo2}) satisfies OT (\ref{bruno5}) with the equality sign, because the scattering is the only source of dissipation. Eq.~(\ref{bo2}) also holds causality. This causality feature is also valid for (\ref{bo}). However, (\ref{bo}) does not generally hold OT (\ref{bruno5}); e.g., because (\ref{bo}) allows for cases with ${\rm Im}[\hat\alpha]<0$. Table~\ref{tab1} shows two of many examples, where (\ref{bo}) fits to the exact resonances in (\ref{reso}).

\begin{table}
\caption{Two examples of matching between the exact resonances (\ref{reso}) and the polarizability (\ref{bo}) of the pumped two-level atom. Both examples refer to ${\rm Im}[\hat\alpha]<0$. The first and second rows refer to the exact resonances $\hat A=1$ and $\hat A+\hat B=-1$, respectively; cf.~(\ref{reso}). }
\begin{tabular}{|c|c|c|c|c|c|}
\hline
$\delta$  & $\epsilon$ & $\om r$ & $|V|$ & ${\rm Im}[\hat{\alpha}]$ & ${\rm Re}[\hat{\alpha}]$ \\
\hline
$1.$ & $0.453379$ & $0.45$ & $1.214541$ & $-0.006337$ & $-0.099312$  \\
\hline
$0.8$ & $0.033625$ & $0.2$ & $1.213356$ & $-0.000010$ & $0.003922$  \\
\hline
\end{tabular}
\label{tab1}
\end{table}

{\it One electric and one magnetic dipoles.}
Given the magnetic polarizability $\chi(\om)$
of the second (magnetic) dipole, we have:
\be
\bvep_{1\, ik}^{ab}
=-\alpha(\om)\delta_{ik}\de^{a1}\de^{b1}, \qquad
\bvep_{2\, ik}^{ab}=-\chi(\om)\delta_{ik}\de^{a2}\de^{b2}.
\label{gloss2}
\ee
We will not write down analogs of (\ref{kosh}), but identify resonances from the poles of the Green function [see SM, section 2]:
\be
\hat \alpha\hat\chi=f(y)\equiv {y^4 e^{-2iy}}{(iy-1)^{-2}}, ~~~ y=\om r,
\label{ono}
\ee
where $\hat \alpha\equiv \frac{\alpha\om^3}{4\pi}$ and $\hat \chi\equiv \frac{\chi\om^3}{4\pi}$ are dimensionless polarizabilities. Now $\hat\alpha$ and $\hat\chi$ holding (\ref{ono}) cannot simultaneously satisfy OT (\ref{bruno5}); see section 2 of SM.

{Both $\hat\alpha$ and $\hat\chi$ cannot hold (\ref{bruno5}), but one of them can}. An interesting case of this is when $\hat\chi$ is small (which is the standard situation with magnetic dipole response \cite{landau}) and holds OT. For clarity, let us assume the same form of $\hat\chi$ as for a two-level resonant atom: $\hat\chi=i3\epsilon/2$, with $\epsilon\ll 1$; cf.~(\ref{bo2}). It satisfies OT (\ref{bruno5}). Now expanding $f(y)$ in (\ref{ono}) for $y<1$ we get $\frac{3i}{2}\epsilon\hat\alpha=y^4(1-y^2)-\frac{2iy^7}{3}+{\cal O}(y^8)$. For $\epsilon$ (small single magnetic response) we can choose $y<1$ so that the latter equation is satisfied with nonequilibrium $\hat\alpha\sim 1$: the influence of the magnetic moment is strongly amplified due to the resonance. Note the difference in our setup and conclusion from Kerker conditions; see \cite{kerker_review} for a review. Those conditions determine the consequences of a similar electric and magnetic moment, while here we consider the amplification of a weak magnetic moment.

\begin{figure}
\includegraphics[width=0.6\columnwidth]{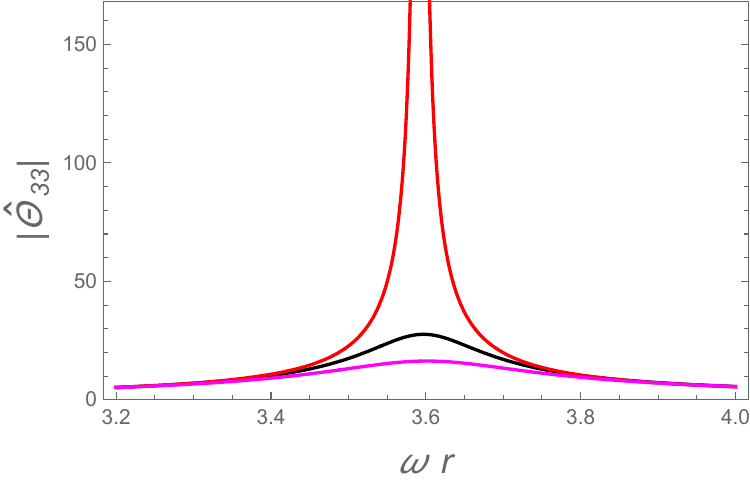}%
  \caption{The absolute value of the effective polarizability $|\hat\Theta_{33}|$ versus $\omega r=\omega|\bm{r}|$ for two gold particles with $\ell=6\times 10^{-5}$ cm holding the Drude model (\ref{cato4}). We set: $\xi=\zeta=1$, $\bm{k}\perp \bm{r}$ and $\hat{\bm{x}}\perp \bm{r}=(r,0,0)$, as in Fig.~\ref{fig1}. For all 3 curves, $\om$ changes in the vicinity of $\gamma$; see (\ref{au}). 
    Red curve: $\om=3.74532\times 10^{13}\,s^{-1}$, $\hat\alpha=-2.73929+i2.52204$. The red curve reaches $\approx 1600$ amplifying $\alpha$ in (\ref{kosh}).   
  Black: $\om=3.57143\times 10^{13}\,s^{-1}$, $\hat\alpha=-2.49781+i2.41167$. 
  Magenta: $\om=3.44828\times 10^{13}\,s^{-1}$, $\hat\alpha=2.32995(-1+i)$.
  The resonance approaches infinity by properly tuning both the inter-particle distance $r$ and $\om$; e.g. $|\hat\Theta_{33}|=1.12\times 10^6$, when (for $\ell=6\times 10^{-6}$ cm) $\omega r=3.59224$ and $\om=3.7484\times 10^{13}\,s^{-1}$. 
    }
  \label{fig2}
\end{figure}

\begin{figure}
\includegraphics[width=0.6\columnwidth]{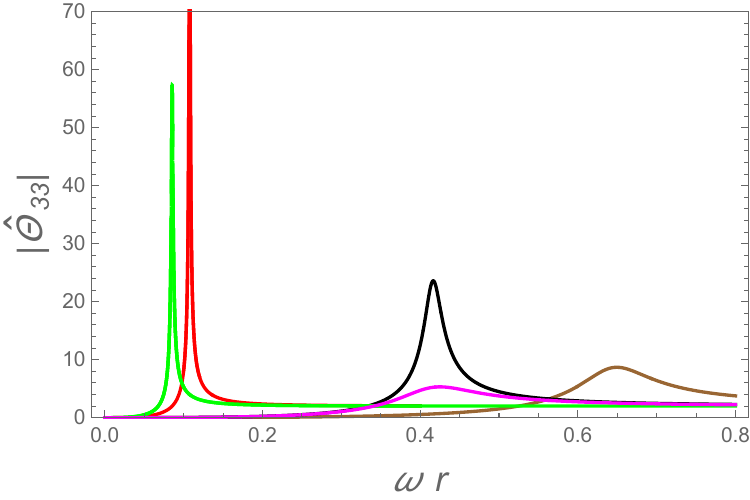}%
  \caption{The absolute value of the effective polarizability $|\hat\Theta_{33}|$ versus $\omega r=\omega|\bm{r}|$. 
  $|\hat\Theta_{33}|$ is defined from (\ref{motik}) with $\xi=\zeta=1$, i.e., $\bm{k}\perp \bm{r}$ and $\hat{\bm{x}}\perp \bm{r}=(r,0,0)$, where $\bm{r}$ is the inter-particle radius-vector, $\hat{\bm{x}}$ is the (far-field) observation direction, and $\bm{k}$ is the wave-vector of the incident wave. The dimensionless polarizability $\hat\alpha$ holds the Drude form (\ref{cato4}) for two identical gold nanoparticles with size $\ell$; see (\ref{au}). 
  Red curve: $\om=2\times 10^{15}\,s^{-1}$, $\ell=10^{-6}$ cm, $\hat\alpha=-1.25\times 10^{-3}+i 2.16\times 10^{-5}$. The red curve reaches $110.7$ amplifying $\approx 10^2$ times $\alpha$ in (\ref{kosh}).
  Green curve: $\om=10^{15}\,s^{-1}$, $\ell=10^{-6}$ cm, $\hat\alpha=-6.25\times 10^{-4}+i 2.15\times 10^{-5}$. 
  Black curve: $\om=10^{15}\,s^{-1}$, $\ell=5\times 10^{-6}$ cm, $\hat\alpha=-7.81\times 10^{-2}+i2.69\times 10^{-3}$.
  Brown curve: $\om=10^{15}\,s^{-1}$, $\ell=8\times 10^{-6}$ cm. 
  Magenta curve: $\om=10^{14}\,s^{-1}$, $\ell=11\times 10^{-6}$ cm. }
  \label{fig1}
\end{figure}

{\it Metallic particles within the Drude model.}
Above, we focused on the exact resonance, where the scattered field in the far-field zone is explicitly infinite. Thus, single-particle polarizability was selected to satisfy the exact resonance condition. Here we consider a more realistic model, where the freedom of selecting parameters is reduced, and hence the resonances are finite. The polarizability of a metallic particle with size $\ell$ [see (\ref{popo})] is due to {quasi-free valence electrons} and is described by the Drude model \cite{klimov} with the dimensionless 
polarizability [cf.~(\ref{on5}, \ref{on6})]:
\BEA
\label{cato4}
\hat\alpha={(\om_{\rm pl}/\om)^2(\om{\ell})^3}
\,[\,(\omega_0/\om)^2-1 -i(\gamma/\omega)]^{-1},
\EEA 
where $\om_{\rm pl}$ is the plasma frequency for the quasi-free electron gas of the particle, $m$ is the electron mass, $1/\gamma$ is the relaxation time due to electron-phonon interactions, and $\omega_0={\hbar}/({m\ell^2})$ \cite{klimov}. 
Parameters of (\ref{cato4}) read for gold (Au) particles \cite{klimov}: 
\be
\label{au}
\om_{\rm pl}=4.11\times 10^{15},\quad \gamma=3.44\times 10^{13}~s^{-1}. 
\ee
Eq.~(\ref{cato4}) holds the causality condition. 
We emphasize that (\ref{cato4}) contains the finite dipole size $\ell$, which was taken to be zero in, e.g.,  (\ref{16}). Thus, condition (\ref{popo}) is essential for applying (\ref{cato4}) to our situation. Eq.~(\ref{cato4}) satisfies OT (\ref{bruno5}) for 
$\gamma\geq{2}\om_{\rm pl}^2\om^2\ell^3/3$, 
which can be realized for a sufficiently small $\ell$. For such cases, the Drude model describes equilibrium physics. We also apply (\ref{cato4}) in a nonequilibrium situation, when (\ref{bruno5}) does not hold. In this context, we emphasize that there is a range of $\ell$, where the OT is not satisfied, but the point-particle limit (\ref{popo}) still applies. 

{\it Maximization of the single-particle scattering.}   
To understand implications of two-particle resonances, we need to maximize the magnitude of the field scattered from a single dipole. It is given by (\ref{kosh}, \ref{cato4}), where we put $\hat\Theta_{ik}=\delta_{ik}$. The maximization of the absolute value of the scattered field over $\om$ and the particle size $\ell$ amounts to maximizing $\frac{1}{4\pi}\om^2|\alpha(\om)|=|\hat\alpha(\om)|/\om$. This maximization is a two-step process. 
First, we maximize $|\hat\alpha(\om)|/\om$ over $\om$ for a fixed $\ell$. When using (\ref{cato4}), this amounts to minimization over $\om$ the quadratic expression 
${[(\omega_0/\om)^2-1]^2 +(\gamma/\omega)^2}$, and produces for the optimal frequency ($\omega_0\equiv {\hbar}/({m\ell^2})$): 
\be
\om_{\rm opt}=\om_0\Big[1-{\gamma^2}/({2\om_0^2})\Big]^{-1/2}\quad 
{\rm for} \quad 1>{\gamma^2}/({2\om_0^2}), \\
\om_{\rm opt}\to\infty\quad {\rm for}\quad 1<{\gamma^2}/({2\om_0^2}).
\label{oho}
\ee
{For sufficiently large $\ell$ we are in the regime (\ref{oho}), which for Au particles is realized for $\ell\gg 0.25\times 10^{-6}$ cm; cf.~(\ref{au}). 
We can take $\ell\simeq 10^{-5}-10^{-4}$ cm for Au. In the regime (\ref{oho}), $\om_{\rm opt}\to\infty$ produces nearly the same result as $\om\lesssim\gamma$. This is the plasmonic resonance of a single metallic particle \cite{klimov}. 
Now the maximization over $\ell$ amounts to maximizing $|\hat\alpha|\propto\ell^3$ from (\ref{cato4}) over $\ell$. Hence, $\ell$ reaches the highest value, which is consistent with point-like condition (\ref{popo}). }

{\it Exact Drude resonances.} 
Figure~\ref{fig2} presents a two-particle scattering regime, where $\hat\alpha$ 
produces the maximal single-particle scattering in the sense defined above. 
Recall that the maximization of the single-particle scattering goes over
$\om$ and $\ell$, still holding (\ref{popo}). Within this regime, OT is not valid. Hence, exact resonances are possible. Fig.~\ref{fig2} captured a vicinity of an exact resonance. The infinite amplification of this resonance can be approached by varying two parameters, e.g., $\om$ and $r$ for a fixed particle size $\ell$.

{{\it Approximate, equilibrium Drude resonances.}}
We now study the two-particle resonances for the Drude model (\ref{cato4}), 
{which do hold OT}. Fig.~\ref{fig1} shows the regime, where $\ell$ is sufficiently small, and OT holds. Sizable, but finite resonances are present. They depend on $\om$ and $\ell$, and amplify (up to $10^2$ times) the single particle response at the same value of $\ell$ and $\om$; see the caption of Fig.~\ref{fig1}. Parameters of Fig.~\ref{fig1} hold $\om\lesssim\gamma$, which is partially optimal for the single-particle response, as discussed above. Fig.~\ref{fig1} shows a narrow resonance (red curve), and smaller resonances obtained for smaller values of $\om$ and larger values of $\ell$. These resonances (black, brown, and magenta curves in Fig.~\ref{fig1}) are smaller and wider, which makes their observation easier. They have larger values of $|{\rm Re}[\hat\alpha]|$ and $|{\rm Im}[\hat\alpha]|$; see the caption of Fig.~\ref{fig1}. Hence, the exact (infinite) resonances -- which violate OT -- are replaced by (finite) approximate resonances, when the theorem is valid.

The maximal value of $|\hat\Theta_{33}|$ in (\ref{kosh}) is 
$>1835.35$. This was found numerically for Au particles, 
given the constraints (\ref{popo}) and (\ref{bruno5}) on (resp.) the point-like limit and the validity of OT; see section 3 of SM for numerical details.

\comment{The analytical description of approximate resonances is found from (\ref{motik}, \ref{kotik}) noting that $\zeta=\xi=1$ for parameters of Fig.~\ref{fig1}. Then we get $\hat\Theta_{33}=2/(1+\hat A)$. For approximate resonances in Fig.~\ref{fig1}, we have ${\rm Re}[\hat\alpha]<0$ and $|{\rm Re}[\hat\alpha]|\gg {\rm Im}[\hat\alpha]$; see the caption of Fig.~\ref{fig1}. Hence, for $y=\om r\ll 1$, we have $1+\hat A\approx 1-|{\rm Re}[\hat\alpha]|\, y^{-3}$, and approximate resonances are present in $\hat\Theta_{33}$. Likewise, we get $\hat\Theta_{11}=2/(1+\hat A+\hat B)$ from (\ref{kotik}), and approximate resonances for $y\ll 1$ are now impossible, since $1+\hat A+\hat B\approx 1+2|{\rm Re}[\hat\alpha]|\,y^{-3}$.}

There is a remote analogy between approximate resonances and the concept of plasmon hybridization reviewed in \cite{klimov}. One clear difference is that plasmon hybridization starts from discrete characteristic values of the single-particle polarizability, which is not needed for approximate resonances. 


{\it Approximate resonances are weaker than the optimal single-particle scattering. }
{We checked that while approximate, equilibrium resonances can improve the partially optimized single-particle scattering, the optimal single-particle scattering can be improved only via violations of OT, i.e., in a nonequilibrium situation; compare Fig.~\ref{fig1} and Fig.~\ref{fig2}. Thus, the gain obtained by the equilibrium two-particle resonances is local in parameter space, enhancing the response compared to a single particle with the same $\omega$ and $l$, but it does not exceed the globally optimized single-particle response.}  
We reached the same conclusion for more general models as well. For improving on the single-particle scattering, we need nonequilibrium resonances, with infinite amplification.

In summary, we have identified a new class of resonances in light scattering from pairs of dipoles. The class originates from multiple scattering and corresponds to poles of the two-particle Green function. These resonances are collective in nature and fully consistent with causality, but, importantly, they can not be realized in equilibrium systems due to OT. While equilibrium systems exhibit only finite amplification, nonequilibrium systems allow access to exact resonances, giving rise to unbounded scattering response. These results reveal a new regime of light scattering, governed by resonant two-particle near-field interaction. They give a pathway toward precision sensing via strong enhancement of weak electromagnetic response; {e.g., this can be applied for measuring weak magnetic fields}.

\acknowledgements

The HESC of Armenia supported us under grants 24FP-1F030, 21AG-1C038 and 22IRF-06. We acknowledge discussions with D. Petrosyan, H. Avetisyan, S. Scheel, and E. Karimi.


\bibliography{references}

\begin{thebibliography}{14}
\expandafter\ifx\csname natexlab\endcsname\relax\def\natexlab#1{#1}\fi
\expandafter\ifx\csname bibnamefont\endcsname\relax
  \def\bibnamefont#1{#1}\fi
\expandafter\ifx\csname bibfnamefont\endcsname\relax
  \def\bibfnamefont#1{#1}\fi
\expandafter\ifx\csname citenamefont\endcsname\relax
  \def\citenamefont#1{#1}\fi
\expandafter\ifx\csname url\endcsname\relax
  \def\url#1{\texttt{#1}}\fi
\expandafter\ifx\csname urlprefix\endcsname\relax\def\urlprefix{URL }\fi
\providecommand{\bibinfo}[2]{#2}
\providecommand{\eprint}[2][]{\url{#2}}

\bibitem[{\citenamefont{Novotny and Hecht}(2012)}]{novotny2012principles}
\bibinfo{author}{\bibfnamefont{L.}~\bibnamefont{Novotny}} \bibnamefont{and}
  \bibinfo{author}{\bibfnamefont{B.}~\bibnamefont{Hecht}},
  \emph{\bibinfo{title}{Principles of nano-optics}}
  (\bibinfo{publisher}{Cambridge university press}, \bibinfo{year}{2012}).

\bibitem[{\citenamefont{Klimov}(2014)}]{klimov}
\bibinfo{author}{\bibfnamefont{V.}~\bibnamefont{Klimov}},
  \emph{\bibinfo{title}{Nanoplasmonics}} (\bibinfo{publisher}{CRC press},
  \bibinfo{year}{2014}).

\bibitem[{\citenamefont{Bohren and Huffman}(2008)}]{bohren2008absorption}
\bibinfo{author}{\bibfnamefont{C.~F.} \bibnamefont{Bohren}} \bibnamefont{and}
  \bibinfo{author}{\bibfnamefont{D.~R.} \bibnamefont{Huffman}},
  \emph{\bibinfo{title}{Absorption and scattering of light by small particles}}
  (\bibinfo{publisher}{John Wiley \& Sons}, \bibinfo{year}{2008}).

\bibitem[{\citenamefont{Mishchenko et~al.}(2006)\citenamefont{Mishchenko,
  Travis, and Lacis}}]{mishchenko2006multiple}
\bibinfo{author}{\bibfnamefont{M.~I.} \bibnamefont{Mishchenko}},
  \bibinfo{author}{\bibfnamefont{L.~D.} \bibnamefont{Travis}},
  \bibnamefont{and} \bibinfo{author}{\bibfnamefont{A.~A.} \bibnamefont{Lacis}},
  \emph{\bibinfo{title}{Multiple scattering of light by particles: radiative
  transfer and coherent backscattering}} (\bibinfo{publisher}{Cambridge
  University Press}, \bibinfo{year}{2006}).

\bibitem[{\citenamefont{Hergert and Wriedt}(2012)}]{mie}
\bibinfo{author}{\bibfnamefont{W.}~\bibnamefont{Hergert}} \bibnamefont{and}
  \bibinfo{author}{\bibfnamefont{T.}~\bibnamefont{Wriedt}},
  \emph{\bibinfo{title}{The Mie theory: basics and applications}}, vol.
  \bibinfo{volume}{169} (\bibinfo{publisher}{Springer}, \bibinfo{year}{2012}).

\bibitem[{\citenamefont{Khanbekyan and Scheel}(2022)}]{khanbekyan}
\bibinfo{author}{\bibfnamefont{M.}~\bibnamefont{Khanbekyan}} \bibnamefont{and}
  \bibinfo{author}{\bibfnamefont{S.}~\bibnamefont{Scheel}},
  \bibinfo{journal}{Physical Review A} \textbf{\bibinfo{volume}{105}},
  \bibinfo{pages}{053711} (\bibinfo{year}{2022}).

\bibitem[{\citenamefont{Markel'}(1992)}]{markel1992scattering}
\bibinfo{author}{\bibfnamefont{V.}~\bibnamefont{Markel'}},
  \bibinfo{journal}{Journal of Modern Optics} \textbf{\bibinfo{volume}{39}},
  \bibinfo{pages}{853} (\bibinfo{year}{1992}).

\bibitem[{\citenamefont{Carney et~al.}(2004)\citenamefont{Carney, Schotland,
  and Wolf}}]{wolf_ot}
\bibinfo{author}{\bibfnamefont{P.~S.} \bibnamefont{Carney}},
  \bibinfo{author}{\bibfnamefont{J.~C.} \bibnamefont{Schotland}},
  \bibnamefont{and} \bibinfo{author}{\bibfnamefont{E.}~\bibnamefont{Wolf}},
  \bibinfo{journal}{Physical Review E} \textbf{\bibinfo{volume}{70}},
  \bibinfo{pages}{036611} (\bibinfo{year}{2004}).

\bibitem[{\citenamefont{Landau and Lifshitz}(2013)}]{landau}
\bibinfo{author}{\bibfnamefont{L.}~\bibnamefont{Landau}} \bibnamefont{and}
  \bibinfo{author}{\bibfnamefont{E.}~\bibnamefont{Lifshitz}},
  \emph{\bibinfo{title}{Electrodynamics of continuous media}},
  vol.~\bibinfo{volume}{8} (\bibinfo{publisher}{elsevier},
  \bibinfo{year}{2013}).

\bibitem[{\citenamefont{Boyd}(2003)}]{boyd}
\bibinfo{author}{\bibfnamefont{R.~W.} \bibnamefont{Boyd}},
  \emph{\bibinfo{title}{Nonlinear Optics}}, vol. \bibinfo{volume}{611}
  (\bibinfo{publisher}{Academic Press}, \bibinfo{year}{2003}).

\bibitem[{\citenamefont{Bettles et~al.}(2016)\citenamefont{Bettles, Gardiner,
  and Adams}}]{dipole_many_2016}
\bibinfo{author}{\bibfnamefont{R.~J.} \bibnamefont{Bettles}},
  \bibinfo{author}{\bibfnamefont{S.~A.} \bibnamefont{Gardiner}},
  \bibnamefont{and} \bibinfo{author}{\bibfnamefont{C.~S.} \bibnamefont{Adams}},
  \bibinfo{journal}{Physical review letters} \textbf{\bibinfo{volume}{116}},
  \bibinfo{pages}{103602} (\bibinfo{year}{2016}).

\bibitem[{\citenamefont{Liu and Kivshar}(2018)}]{kerker_review}
\bibinfo{author}{\bibfnamefont{W.}~\bibnamefont{Liu}} \bibnamefont{and}
  \bibinfo{author}{\bibfnamefont{Y.~S.} \bibnamefont{Kivshar}},
  \bibinfo{journal}{Optics express} \textbf{\bibinfo{volume}{26}},
  \bibinfo{pages}{13085} (\bibinfo{year}{2018}).

\bibitem[{\citenamefont{Ballentine}(2014)}]{ballentine}
\bibinfo{author}{\bibfnamefont{L.~E.} \bibnamefont{Ballentine}},
  \emph{\bibinfo{title}{Quantum mechanics: a modern development}}
  (\bibinfo{publisher}{World Scientific Publishing Company},
  \bibinfo{year}{2014}).

\bibitem[{\citenamefont{Nieuwenhuizen et~al.}(1992)\citenamefont{Nieuwenhuizen,
  Lagendijk, and van Tiggelen}}]{theo}
\bibinfo{author}{\bibfnamefont{T.~M.} \bibnamefont{Nieuwenhuizen}},
  \bibinfo{author}{\bibfnamefont{A.}~\bibnamefont{Lagendijk}},
  \bibnamefont{and} \bibinfo{author}{\bibfnamefont{B.~A.} \bibnamefont{van
  Tiggelen}}, \bibinfo{journal}{Physics Letters A}
  \textbf{\bibinfo{volume}{169}}, \bibinfo{pages}{191} (\bibinfo{year}{1992}).

\end{thebibliography}

\section*{Supplementary Material}

This supplementary material discusses the derivation of our main results via the Lippmann-Schwinger equation (section 1). Section 2 presents derivation details for one electric and one magnetic dipole.
Section 3 discusses the maximal amplification of approximate Drude resonances.

{\bf 1. Lippmann-Schwinger equations for point scatterers. }

We aim to study scattering from point-like dipoles. Consider a monochromatic electromagnetic field ($k=1,2,3$):
\be
E_k(\bm{x},t)=e^{-i\om t}E_k(\bm{x},\om),~~ 
H_k(\bm{x},t)=e^{-i\om t}H_k(\bm{x},\om). 
\label{ii2}
\ee
The two Maxwell's equations for 
$E_k(\bm{x},\om)=\E^1_k$ and $H_k(\bx,\om)=\E^2_k$ read in Gaussian units and in a representation that combines spinors (denoted by upper indices $a,b=1,2$) and 3-dimensional vectors (lower indices $i,j,k=1,2,3$) \cite{landau}: 
\be
\label{x22}
&M\begin{bmatrix}
\E^1_k\\
\E^2_k
\end{bmatrix}=
\eta
\begin{bmatrix}
\E^1_k\\
\E^2_k
\end{bmatrix},  ~~~~
M\equiv \begin{bmatrix}
\delta_{ik} & \frac{1}{i\om}\ep_{ijk}\pa_j \\
\frac{-1}{i\om}\ep_{ijk}\pa_j  & \delta_{ik} 
\end{bmatrix},
\\
&
\eta_{ik}^{ab}(\bx)=[1-\vep(\bx,\om)]\delta_{ik}\delta^{a1}\delta^{b1}
+[1-\mu(\bx,\om)]\delta_{ik}\delta^{a2}\delta^{b2},
\ee
where $c=1$, we imply summation over repeated indices, $\delta_{ik}$ is the Kroenecker symbol, $\bx=(x_1,x_2,x_3)$, $\pa_i=\pa_{x_i}$, and $\ep_{ijk}$ is the fully antisymmetric tensor: $\ep_{123}=1=-\ep_{213}$. $\vep(\bx,\om)$ and $\mu(\bx,\om)$ are scalar space-dependent scalar electric permittivity and magnetic permeability, which characterize the scattering medium. 
Free Green's functions of (\ref{x22}) are found from 
\begin{align}
M\begin{bmatrix}
g^{11}_{kl}(\bx) & g^{12}_{kl}(\bx)\\
g^{21}_{kl}(\bx) & g^{22}_{kl}(\bx)\\
\end{bmatrix}= I_{il}\delta(\bx),\quad I_{il}\equiv
\delta_{il}
\begin{bmatrix}
1 & 0\\
0 & 1\\
\end{bmatrix},
\label{gre}
\end{align}
which are solved via the scalar free Green function $g_0(x)$
\begin{align}
\label{61}
&(\pa_i\pa_i+\om^2)g_0(x)=\de(\bx),~~~ g_0(x)=-\frac{e^{i\om x}}{4\pi x},~~~ x=|\bx|,\\
\label{62}
&g^{11}_{il}(\bx)=g^{22}_{il}(\bx)=\de_{il}\om^2 g_0(x)+\pa^2_{il}g_0(x),\\
&g^{21}_{il}(\bx)=-g^{12}_{il}(\bx)=i\om \ep_{ils}\pa_sg_0(x),
\label{e1}
\end{align}
where in (\ref{61}) we choose the retarded Green functions as needed for scattering  \cite{ballentine}. The electromagnetic field in (\ref{x22}) reads
\be
F(\bx)=F^{[\rm in]}(\bx)+F^{[\rm sc]}(\bx),\quad MF^{[\rm in]}(\bx)=0,
\ee
where the incident field $F^{[\rm in]}$ holds free equations (\ref{x22}), while $F^{[\rm sc]}$ is the scattered field. We focus on $n=1,2$ point dipole scatterers:
\be
\label{162}
\bvep(\bx)={\sum}_{\sigma=1}^n\bvep_\sigma\delta(\bx-\ba_\sigma),\qquad 
\bvep_\sigma=\{\bvep_{\sigma,\,ik}^{ab}\}, 
\ee
where $\quad i,k=1,2,3$; $a,b=1,2$, and $\bvep_\sigma$ is the electromagnetic polarizability of the dipole located at $\ba_\sigma$.
Eq.~(\ref{162}) implies that both the wavelength $1/\om$ and inter-particle distances 
$|\ba_\sigma-\ba_\mu|$ are much larger than the dipole particle's size:
\be
{1}/{\om},\,\,  |\ba_\sigma-\ba_\mu| \gg \ell\equiv {\rm dipole~size}.
\label{popo2}
\ee
We get from (\ref{x22}--\ref{162}) the Lippmann-Schwinger equations for the full Green function $G=(M-\eta)^{-1}$, which determines  $\E^{\rm [sc]}$ \cite{ballentine}:
\begin{align}
\label{o1}
&G(\bx,\ba_\nu)=g(\bx-\ba_\nu)+
{\sum}_{\sigma=1}^ng(\bx-\ba_\sigma)\eta_\sigma G(\ba_\sigma,\ba_\nu),\\
&\E^{\rm [sc]}(\bx)={\sum}_{\nu=1}^n
G(\bx,\ba_\nu)\eta_\nu \E^{\rm [in]}(\ba_\nu).
\label{o11}
\end{align}
The limit (\ref{162}, \ref{popo2}) makes the scattering problem solvable. But it involves infinities and needs renormalization \cite{theo}. We start with one point scatterer located at $\ba_1=0$. We get from (\ref{o1}):
$G(\bx,0)=g(\bx)+g(\bx)\bvep_1 G(0,0)$, 
$G(0,0)=[I-g(0)\bvep_1]^{-1}g(0)$.
These contain infinite expressions $g(0)=\infty$ (due to $g_0(0)=\infty$ in (\ref{61})). The renormalization program for model (\ref{162}) amounts to taking $g(0)$ finite
\cite{theo}, and expressing $F^{\rm [sc]}$ via single-particle polarizabilities $\eta_\sigma$ so that no new phenomenological parameters are generated. This program can be carried out for $n$ point-like dipoles in (\ref{162}). A simple shortcut to the full renormalizability is to assume $g(0)=0$, where the self-scattering is excluded. Due to $G(0,0)=0$, we get: $\E^{\rm [sc]}(\bx)=g(\bx)\bvep_1 \E^{\rm [in]}(0)$.

Consider in (\ref{162}) two point scatterers located at $\ba_1$ and $\ba_2$: 
$\bm{r}=\ba_1-\ba_2$, $r=|\bm{r}|$.
Using (\ref{o1}) and $g(0)=0$ we find in block-matrix notations 
\begin{align}
\label{175}
\begin{bmatrix}
I & -g(\bm{r})\bvep_2 \\
-g(-\bm{r})\bvep_1 & I\\
\end{bmatrix}
\begin{bmatrix}
G(\ba_1,\by)\\
G(\ba_2,\by)\\
\end{bmatrix}
=
\begin{bmatrix}
g(\ba_1-\by)\\
g(\ba_2-\by)\\
\end{bmatrix}.
\end{align}
We get from inverting (\ref{175})
\be
\label{20}
\begin{bmatrix}
G(\ba_1,\by)\\
G(\ba_2,\by)\\
\end{bmatrix}
=\begin{bmatrix}
h_1& g(\bm{r})\bvep_2h_2\\
g(-\bm{r})\bvep_1h_1  & h_2 \\
\end{bmatrix}
\begin{bmatrix}
 g(\ba_1-\by)\\
 g(\ba_2-\by)\\
\end{bmatrix},
\ee
where $h_1$ and $h_2$ depend on inter-dipole distance, $\eta_1$ and $\eta_2$:
\be
\label{adyge1}
h_1=[I-g(\bm{r})\bvep_2 g(-\bm{r})\bvep_1]^{-1},\\
h_2=[I- g(-\bm{r})\bvep_1 g(\bm{r})\bvep_2]^{-1}, 
\label{adyge2}
\ee
$h_1$ and $h_2$ come from the multiple scattering of light from one dipole to another, as seen from the Green functions
$g(-\bm{r})$and $g(\bm{r})$. We confirm below that the resonances are zero eigenvalues of $h_1^{-1}$ and $h_2^{-1}$ in (\ref{adyge1}, \ref{adyge2}). Eqs.~(\ref{o11}, \ref{20}) determine the scattered field, where $G(\ba_1,\ba_1)$, $G(\ba_2,\ba_2)$, $G(\ba_1,\ba_2)$, and $G(\ba_2,\ba_1)$
are found from (\ref{20}) and $g(0)=0$.

We make two assumptions in (\ref{o11}). First, we work in the far-field limit, where $|\bx|\gg|\ba|$ and $g(\bx-\ba)$ is approximated from (\ref{61}--\ref{e1}) as ($\hat \bx=\bx/|\bx|$ and $x=|\bx|$):
\be
g(\bx-\ba)\simeq -\om^2\,\frac{e^{i\om x-i\ba\cdot\hat\bx}}{4\pi x}
\begin{bmatrix}
\de_{kl}-\hat x_k\hat x_l& \ep_{kls}\hat x_s\\
-\ep_{kls}\hat x_s & \de_{kl}-\hat x_k\hat x_l \\
\end{bmatrix}, 
\label{far}
\ee
Second, we assume that the incident field is a plane wave 
\be
\label{31}
\E^{\rm [in]}(\ba_1)=e^{i\bm{k}\ba_1}\E^{\rm [in]}(0), ~~
\E^{\rm [in]}(\ba_2)=e^{i\bm{k}\ba_2}\E^{\rm [in]}(0), 
\ee
where $\bk$ is the wave-vector: $|\bk|=\om$. These assumptions are natural for observing the resonances, though they amplify any incident field, both in far field and near-field domains; see below.

For two identical electric dipoles, we have in (\ref{162}):
$\bvep_{1\, ik}^{ab}(\om)=\bvep_{2\, ik}^{ab}(\om)
=-\alpha(\om)\delta_{ik}\de^{a1}\de^{b1}$, 
where $\alpha(\om)/(4\pi)$ is the dipole polarizability. 
We get from (\ref{o11}, \ref{far}, \ref{31}) for the scattered field 
\be
\label{3332}
\begin{bmatrix}
E^{\rm [sc]}_k(\bx) \\
H^{\rm [sc]}_k(\bx)
\end{bmatrix}
=-
\begin{bmatrix}
(\de_{kl}-\hat x_k\hat x_l) \hat\Theta_{ln} E_n^{\rm [in]}(0) \\
-\ep_{kls}\hat x_s \hat\Theta_{ln} E_n^{\rm [in]}(0) \\
\end{bmatrix}\times \nonumber\\
\alpha \om^2 g_0(x) e^{i\ba_1(\bk-\om\hat\bx)}.
\label{kosh2}
\ee

{\bf 2. One electric and one magnetic dipole}

Given the magnetic polarizability $\chi(\om)$
of the second (magnetic) dipole, we have:
\be
\bvep_{1\, ik}^{ab}
=-\alpha(\om)\delta_{ik}\de^{a1}\de^{b1}, \qquad
\bvep_{2\, ik}^{ab}=-\chi(\om)\delta_{ik}\de^{a2}\de^{b2}.
\label{gloss22}
\ee
We will not write down analogs of (\ref{kosh2}), but employ (\ref{gloss22}) to identify resonances from zero eigenvalues of $h_1^{-1}$ and $h_2^{-1}$ in (\ref{adyge1}, \ref{adyge2}); cf.~the remark after (\ref{o1}) and see (\ref{e1}). We get that the resonances are found from [see (\ref{61}--\ref{e1})]: 
\be
\alpha\chi\om^2 g_0'(r)^2=1,
\ee
which can also be written as 
\be
\hat \alpha\hat\chi=f(y)\equiv {y^4 e^{-2iy}}{(iy-1)^{-2}}, ~~~ y=\om r,
\label{ono2}
\ee
where $\hat \alpha\equiv \frac{\alpha\om^3}{4\pi}$ and $\hat \chi\equiv \frac{\chi\om^3}{4\pi}$ are dimensionless polarizabilities. Let us now show that $\hat\alpha$ and $\hat\chi$ holding (\ref{ono2}) cannot simultaneously satisfy the optical theorem (OT) 
\BEA
{\rm Im}[\hat \alpha]\geq 2|\hat \alpha|^2/3,\qquad
{\rm Im}[\hat \chi]\geq 2|\hat \chi|^2/3.
\label{bruno52}
\EEA
To this end, introduce $\hat\alpha=|\hat\alpha|e^{i\theta_1}$, $\hat\chi=|\hat\chi|e^{i\theta_2}$, $f=|f|e^{i\theta_f}$, and multiply together the individual OTs (\ref{bruno52}) for $\hat\alpha$ and $\hat\chi$: 
\be
\label{opo}
\sin\theta_1\sin\theta_2=(\cos[\theta_f-2\theta_1]-\cos[\theta_f])/2\geq{4|f|}/{9},
\ee
where we again used (\ref{ono2}). Maximizing the left-hand side of (\ref{opo}) over $\theta_1$ we reduce (\ref{opo}) to $1-\frac{{\rm Re}[f]}{|f|}\geq\frac{8|f|}{9}$. This inequality cannot hold, as seen from (\ref{ono2}). This method of showing the invalidity of OT at the strong resonance applies more generally, e.g., to two dipoles with different electric polarizabilities.

{\bf 3. The maximal value of $|\hat\Theta_{33}|$.}

What is the maximal value of $|\hat\Theta_{33}|$ in (\ref{kosh2}) given the constraints (\ref{popo2}) and (\ref{bruno52}) on (resp.) the point-like limit and the validity of OT? Numerical maximization over the involved parameters for two identical Au nanoparticles 
produced ${\rm max}\Big[|\hat\Theta_{33}|\Big]>1835.35$, i.e., $\sim 10^3$-times amplification, which is reached for $\om=3.18067\times 10^{16}$ Hz, $\ell=0.7444\times 10^{-7}$ cm, $\om r=0.02017$, $k_x/|\bm{k}|=0.326941$, $\hat x_1=-0.931752$. This amplification requires fine-tuning of parameters.

\end{document}